\documentclass[aps,twocolumn,showpacs]{revtex4}
\usepackage{mathrsfs}
\begin{document}
\title{Testing the correctness of the Poynting vector
$\vec E\times\vec B$ as the momentum density of gauge fields}
\author{Xiang-Song Chen\footnote{Email: cxs@scu.edu.cn}, Xiang-Bai Chen}
\affiliation{Department of Physics, Sichuan University, Chengdu
610064, China}
\author{Wei-Min Sun, Fan Wang}
\affiliation{Department of Physics, Nanjing University, Nanjing
210093, China}
\date{October 8, 2007}

\begin{abstract}
Following our recent finding that the renowned formula $\vec x\times
(\vec E\times\vec B)$ is not the correct density for the
electromagnetic angular momentum,\cite{Chen07a}  here we examine the
validity of the Poynting vector $\vec E\times\vec B$ as the
electromagnetic momentum density (or energy flux). The competitor is
the gauge-invariant canonical momentum $E^i\vec \nabla A^i_\perp$.
It often gives the same result as $\vec E\times\vec B$, but we
propose that a delicate measurement (of the {\em azimuthal} energy
flow in polarized atomic radiations) can make a discrimination. By
clarifying the profound difference between two kinds of
energy-momentum tensors: the canonical (or mechanical) one and the
symmetric (or gravitational) one, we predict that it is $E^i\vec
\nabla A^i_\perp$ that would pass the delicate experimental test.
Our observations have far-reaching implications for understanding
the source of gravity, and the nucleon momentum as well.

\pacs{11.15.-q, 12.20.-m, 32.30.-r, 42.25.Ja}
\end{abstract}

\maketitle

For nearly 100 years, the Poynting vector $\vec E\times\vec B$ was
taken as the momentum (or energy flux) density of the
electromagnetic field. Its correctness has scarcely, if not never,
been doubted. The firm conviction comes from both theory and
experiment. Theoretically, it is a standard textbook exercise to
derive $\vec E\times\vec B$ from energy or momentum conservation
considerations \cite{Jack99}, or from the general definition of
energy-momentum tensor in gravitational theory \cite{Wein72}.
Experimentally, $\vec E\times\vec B$ seems to give satisfactory
descriptions for all phenomena of electromagnetic momentum or energy
flow, even the angular distribution in multipole radiation.

The hints for us to question the validity of $\vec E\times\vec B$
are of 3-fold. Firstly, we have noted long ago that when the gauge
field interacts with the Dirac field, $\int d^3x \vec E\times\vec B$
is no longer its generator of spatial translation, thus does not
qualify for a momentum definition.\cite{Chen97} Secondly, the
angular momentum density built from the Poynting vector, $\vec
x\times (\vec E\times\vec B )$, was recently shown by us to be
incorrect.\cite{Chen07a} The third hint comes from a special
perspective of \emph{understanding momentum from angular momentum}:
If $\vec P(x)$ is the momentum density, then
\begin{equation}
\int d^3x ~ \vec x\times \vec P(x) \label{5}
\end{equation}
is the standard form of orbital
angular momentum. However, we know that the integration
\begin{equation}
\vec J_\gamma\equiv\int d^3x ~\vec x\times (\vec E\times\vec B)
\label{10}
\end{equation}
gives the total angular momentum for a free electromagnetic field,
including both spin and orbital contributions. This implies that
$\vec E\times\vec B$ is not a pure mechanical momentum. Instead, it
includes the spin flow. Such character is already known from the
Gordon decomposition of the electron current into a ``convection''
current plus a spin current:\cite{Bjor64}
\begin{equation}
\vec j^\mu\equiv \bar\psi \gamma^\mu\psi=\frac{1}{2m}\bar\psi \frac
1i (\partial^\mu
-\overleftarrow{\partial^\mu})\psi+\frac{1}{2m}\partial_\nu
(\bar\psi \sigma^{\mu\nu}\psi ). \label{20}
\end{equation}
In consequence the electron magnetic moment $\int d^3 x ~\vec
x\times \vec j$ can be decomposed into an orbital part plus a spin
part.\cite{Chen04} The Poynting vector $\vec E\times\vec B$ can be
decomposed in a similar manner. By writing $\vec B= \vec\nabla
\times \vec A_\perp$ (here $\vec A_\perp$ is the transverse part of
$\vec A$ defined by $\vec\nabla\cdot\vec A_\perp=0$), and using
$\vec \nabla\cdot \vec E=0$ for a free field, we have
\begin{equation}
\vec E\times \vec B=\vec E^i\vec \nabla A^i_\perp +\nabla^i (E^i\vec
A_\perp). \label{30}
\end{equation}
Inserting this decomposition into Eq. (\ref{10}) and integrating by
parts, one gets the separate, gauge-invariant spin and orbital
angular momentum of electromagnetic field:\cite{Enk94}
\begin{equation}
\vec J_\gamma =\int d^3x \vec E\times \vec A_\perp + \int d^3x ~\vec
x\times E^i\vec \nabla A^i_\perp. \label{40}
\end{equation}
Referring to the form of orbital angular momentum in Eq. (\ref{5}),
this suggests that it is the canonical expression $E^i\vec \nabla
A^i_\perp$ that gives the pure mechanical momentum density of
electromagnetic field. Moreover, Eqs. (\ref{10}) and (\ref{40})
immediately suggest that $\vec E\times\vec B$ and $E^i\vec \nabla
A^i_\perp$ can be discriminated by measuring the {\em azimuthal}
momentum (or energy flow), which is responsible for the angular
momentum of the emitted photon in polarized radiations. As long as
the photon spin contributes, then $\vec E\times\vec B$ and $E^i\vec
\nabla A^i_\perp$ must not always have the same azimuthal
components. In the following, we calculate this measurable effect
explicitly.

As in our recent study of the angular momentum
density,\cite{Chen07a} here we again look at a pure multipole
radiation of order $(l,m)$, which provides the most convenient and
well-defined electromagnetic configuration with rich spatial
structure. To make the spin effect relatively significant, we
consider the lowest $(l,m)=(1,1)$, namely, the dipole radiation,
which also has the strongest radiation intensity for easier
measurement. The electric dipole radiation field is given by
\begin{eqnarray}
\vec B&=&\mathscr{A} h^{(1)}_1(kr) \vec L Y_{11},\nonumber \\
\vec E&=&i k \vec A_\perp =\frac ik \vec \nabla \times \vec B.
\label{50}
\end{eqnarray}
The magnetic dipole radiation field is obtained by the interchange
$\vec E\to \vec B$ and $\vec B\to -\vec E$. Here $\vec L\equiv \vec
x\times \frac 1i \vec \nabla$. The wave amplitude $\mathscr{A}$ is
related to the emission probability, which in turn is determined by
the transition matrix element of the electric (magnetic) dipole
moment.

Straightforward calculations give the time-averaged ``energy'' flow
density:
\begin{eqnarray}
\frac 12\mathrm{Re}[\vec E^*\times \vec B]&=&\vec n_r
\frac{\left|\mathscr{A}\right|^2 }{(kr)^2} \cdot\frac{3}{16\pi}\cdot
\frac 12(1+\cos ^2\theta)\nonumber
\\
&+&\vec n_\phi \frac{\left|\mathscr{A}\right|^2}{(kr)^3}
\cdot\frac{3}{16\pi}\cdot \sin\theta,
\label{60}\\
\frac 12\mathrm{Re}[E^{*i}\vec \nabla A^i_\perp]&=&\vec n_r
\frac{\left|\mathscr{A}\right|^2}{(kr)^2} \cdot\frac{3}{16\pi} \cdot
\frac 12(1+\cos ^2\theta)\nonumber
\\
&+&\vec n_\phi\frac{\left|\mathscr{A}\right|^2}{(kr)^3}
\cdot\frac{3}{16\pi}\cdot\frac 12 \sin\theta .\label{61}
\end{eqnarray}
These formulae hold for both electric and magnetic dipole
radiations. Irrelevant higher order terms have been omitted. $\vec
n_r$ and $\vec n_\phi$ are the unit vectors in the radial and
azimuthal directions, respectively. We see that the radial flow
given by $\vec E\times\vec B$ and $E^i\vec \nabla A^i_\perp$ is
exactly the same in the leading order, because at this order $\vec
A_\perp$ of a radiation field has no radial component, thus the
second term in Eq. (\ref{30}) drops out. (This partially explains
how $\vec E\times\vec B$ survived a century.) On the other hand, the
azimuthal flow given by $E^i\vec \nabla A^i_\perp$ is half of that
given by $\vec E\times\vec B$. (From our above analysis, the other
half of $\vec E\times\vec B$ is a spin flow.) This is the effect we
aimed at: \emph{Measurement of the azimuthal energy flow in
polarized atomic radiation can unambiguously tell whether $\vec
E\times\vec B$ or $E^i\vec \nabla A^i_\perp$ is the correct
electromagnetic momentum density, or energy flux.} The intensity
ratio of the azimuthal to radial energy flows is of the order
$(kr)^{-1}$, which would be the major challenge to the measurement.

It indeed sounds unbelievable that in the era when people seek
physics beyond the Standard Model, there still exists
misunderstanding of such most fundamental quantities as the
electromagnetic energy and momentum. The essential ignorance in this
regard during the past century is the profound difference between
two kinds of energy-momentum tensors: the canonical or mechanical
one (hereafter denoted as $T^{\mu\nu}$) and the symmetric or
gravitational one (hereafter denoted as $\Theta^{\mu\nu})$. The
Poynting vector $\vec E\times\vec B$ is a component of the latter,
while $E^i\vec \nabla A^i_\perp$ is a component of the former. These
two tensors are often regarded as being physically identical, for
they just differ by total derivative terms, thus give the same
conserved total energy and momentum. We must remind here that such
identification is indeed rather unthoughtful, because, after all,
the total derivative term does lead to change of the density, which
would inevitably manifest somewhere.

Gravitational theory provides a general definition of a symmetric
$\Theta^{\mu\nu}$ as the ``functional derivative'' of the matter
action $I_M$ with respect to the metric tensor
$g_{\mu\nu}$:\cite{Wein72}
\begin{equation}
\delta I_M\equiv \frac 12 \int d^4x
\sqrt{-g(x)}\Theta^{\mu\nu}(x)\delta g_{\mu\nu}(x).
\end{equation}
The canonical energy-momentum tensor $T^{\mu\nu}$ is directly
derived from the translational invariance of the Lagrangian
$\mathcal{L}=\int d^3 x \mathscr{L} (\phi_i,\partial_\mu\phi_i)$:
\begin{equation}
T^{\mu\nu}=\frac{\partial
\mathscr{L}}{\partial(\partial_\mu\phi_i)}\partial^\nu\phi_i
-g^{\mu\nu}\mathscr{L}.
\end{equation}
[It might be useful to add a remark (mainly for the student reader)
that there exists a third method of deriving the energy-momentum
tensor, by purely mechanical analysis, as is done in textbooks on
classical electrodynamics. This method, however, permits too much
arbitrariness in defining a conserved quantity.]

The distinction between $T^{\mu\nu}$ and $\Theta^{\mu\nu}$ can be
most clearly seen and clarified from the angular momentum
constructed from them:
\begin{eqnarray}
J^{ij}&\equiv& \int d^3 x
(x^i\Theta^{0j}-x^j\Theta^{0i}),\label{70} \\
L^{ij}&\equiv& \int d^3 x (x^iT^{0j}-x^j T^{0i}).\label{71}
\end{eqnarray}
$J^{ij}$ is the conserved total angular momentum of the system. It
agrees with the total canonical angular momentum derived from the
rotational invariance of the Lagrangian. In comparison, $L^{ij}$ is
merely the orbital angular momentum, and is not separately
conserved. Following our perspective of ``understanding momentum
from angular momentum'', we conclude that the mechanical momentum
density should be $T^{0i}$, not $\Theta^{0i}$. The latter must
include a spin current so as to produce the total angular momentum
by an apparent orbital expression in Eq. (\ref{70}). On the other
hand, the non-symmetric $T^{\mu\nu}$ does not fit into Einstein's
gravitational equations, which necessarily require the symmetric
energy-momentum tensor $\Theta^{\mu\nu}$. This reveals a profound
fact that the source of gravitational field is not the pure
energy-momentum. {\em Spin also plays a role in generating gravity!}
Then, a highly serious question may naturally come to the mind:
\emph{Since $T^{\mu\nu}$ and $\Theta^{\mu\nu}$ are fundamentally
different, would Einstein's Equivalence Principle be strict?}

Before closing the paper, we should comment on the gauge invariance
of the canonical momentum. For a free electromagnetic field, the
canonical momentum density $E^i\vec\nabla A^i_\perp$ is gauge
invariant, because the transverse component $\vec A_\perp$ is not
affected by gauge transformations. In our recent discussions of
angular momentum in gauge theories, we showed that gauge-invariant
canonical operators can as well be constructed in an interacting
system of QED or QCD.\cite{Chen07a,Chen07b} We record here the
momentum expression:
\begin{equation}
\vec P=\int d^3x (\psi ^\dagger \frac 1i \vec D_{pure} \psi +
E^i\vec \nabla A^i_{phys}). \label{80}
\end{equation}
Here the ``pure-gauge'' covariant derivative $\vec D_{pure}$ is
$(\vec \nabla -ie\vec A_{pure})$ in QED and $(\vec \nabla -ig\vec
A^a_{pure}T^a)$ in QCD. In both QED and QCD, $\vec A_{pure}$ and
$\vec A_{phys}$ can be consistently defined to be the pure-gauge and
gauge-invariant/covariant components of $\vec A$,
respectively.\cite{Chen07a,Chen07b} $\vec A_{phys}$ can thus be
termed the ``physical'' component. For QED, $\vec A_{pure}$ and
$\vec A_{phys}$ are just the longitudinal and transverse components
of $\vec A$. For QCD, the definitions of $\vec A_{pure}$ and $\vec
A_{phys}$ are not trivial due to non-Abelian features. Details can
be found in \cite{Chen07b}.

In both QED and QCD, there exists a ``physical'' gauge in which the
pure-gauge term vanishes,\cite{Chen07a,Chen07b} then Eq. (\ref{80})
reduces to a naive expression:
\begin{equation}
\vec P=\int d^3x (\psi ^\dagger \frac 1i \vec \nabla \psi + E^i\vec
\nabla A^i). \label{90}
\end{equation}
For QED the ``physical'' gauge is just the Coulomb gauge $\vec
\nabla\cdot\vec A=0$. For QCD it is $[\vec A^aT^a,\vec
E^bT^b]=0$.\cite{Chen07b} Within the ``physical'' gauge, one can use
the naive canonical expression in Eq. (\ref{90}), which would
automatically produce the gauge-invariant quantity defined by Eq.
(\ref{80}).

It is worthwhile to remark that in previous studies of the nucleon
momentum structure, the gluon momentum was taken as $\int d^3 x \vec
E^a\times \vec B^a$, which leads to a picture that gluons carry half
of the nucleon momentum on the light-cone.\cite{Sloa88} From our
discussions here, this picture might need to be revised.

To summarize, we explained that the canonical and symmetric
energy-momentum tensors have fundamental difference in their
physical contents: The former gives the pure mechanical
energy-momentum, while the latter actually includes the spin
current. It should be the canonical instead of symmetric
energy-momentum tensor that can properly describe the energy flow in
experiments. On the other hand, the source for gravitational field
is necessarily the symmetric energy-momentum tensor, which involves
the spin effect. We encourage experimentalists to perform the
measurement we proposed here, so as to clarify the understanding of
such most fundamental conceptions as energy and momentum. We do not
doubt that the experimental findings would confirm our clarification
about the distinct physical contents of the mechanical and symmetric
energy-momentum tensors.

We are grateful to T. Goldman and X.F. L\"u for fruitful
discussions. This research is supported by the National Science
Foundation of China under grants 10475057 and 90503011.

\end{document}